\documentclass[11pt]{article}
\usepackage{fullpage}
\usepackage{amsmath, amsfonts, amsthm, epsf}
\newcommand{\IGNORE}[1]{}

\title{ 
Transdichotomous Results in Computational Geometry, II:\\
Offline Search%
\thanks{A preliminary version of this work with the title
``Voronoi Diagrams in $n\cdot 2^{O(\sqrt{\lg\lg n})}$ Time'' appeared in
{\em Proc.\ 39th ACM Symposium on Theory of Computing}, pages 31--39, 2007.}
}
\author{Timothy M. Chan%
  \thanks{School of Computer Science,
    University of Waterloo,
    Waterloo, Ontario N2L 3G1, Canada
    (tmchan@uwaterloo.ca).  This work has been supported by an NSERC grant.}
  \and Mihai P\v{a}tra\c{s}cu%
  \thanks{AT\&T Labs, Florham Park NJ, USA (mip@alum.mit.edu).
    Part of this work was done while the author was at MIT.
  }
}

\IGNORE{
\documentclass{sig-alternate}
\title{ Voronoi Diagrams in {\LARGE $n\cdot 2^{O(\sqrt{\lg\lg n})}$} Time }
\numberofauthors{2}
\author{\alignauthor
  Timothy M. Chan \\[4pt]
  \affaddr{School of Computer Science}\\
  \affaddr{University of Waterloo}\\
  \affaddr{Waterloo, Ontario N2L 3G1, Canada}\\[2pt]
  \email{tmchan@uwaterloo.ca}
\alignauthor
  Mihai P\v{a}tra\c{s}cu \\[4pt]
  \affaddr{Computer Science \& Artificial Intelligence Lab}\\
  \affaddr{Massachusetts Institute of Technology}\\
  \affaddr{Cambridge, MA 02139, USA}\\[2pt]
  \email{mip@mit.edu} }

\date{}
\conferenceinfo{STOC'07,} {June 11--13, 2007, San Diego, California, USA.}
\CopyrightYear{2007}
\crdata{978-1-59593-631-8/07/0006}
}

\let\latexcite=\cite
\def\cite{\nolinebreak\latexcite}
\let\latexref=\ref
\def\ref{\nolinebreak\latexref}

\newenvironment{enumerate*}%
  {\vspace{-1.3ex}\begin{enumerate}%
      \setlength{\itemsep}{-.2ex}\setlength{\parsep}{-.2ex}}%
  {\vspace{-1ex}\end{enumerate}}

\newenvironment{itemize*}%
  {\vspace{-1.3ex}\begin{itemize}%
      \setlength{\itemsep}{-.2ex}\setlength{\parsep}{-.2ex}}%
  {\vspace{-1ex}\end{itemize}}

\newtheorem{lemma}{Lemma}  

\newtheorem{corollary}[lemma]{Corollary}
\newtheorem{observation}[lemma]{Observation}

\newenvironment{pf}{\noindent{\bf Proof:} }{\hspace*{\fill}$\Box$}
\newenvironment{algm}{\begin{quote}\begin{tabbing}%
  19.M\= for\= for\= for\= for\= for\=\+\kill\<}{\end{tabbing}\end{quote}}
\newcommand{\down}[1]{{\left\lfloor{#1}\right\rfloor}}
\newcommand{\TO}{,\ldots,}

\newcommand{\eps}{\varepsilon}

{\makeatletter
 \gdef\xxxmark{%
   \expandafter\ifx\csname @mpargs\endcsname\relax 
     \expandafter\ifx\csname @captype\endcsname\relax 
       \marginpar{xxx}
     \else
       xxx 
     \fi
   \else
     xxx 
   \fi}
 \gdef\xxx{\@ifnextchar[\xxx@lab\xxx@nolab}
 \long\gdef\xxx@lab[#1]#2{{\bf [\xxxmark #2 ---{\sc #1}]}}
 \long\gdef\xxx@nolab#1{{\bf [\xxxmark #1]}}
}

\newenvironment{myitemize}{\begin{itemize}}{\end{itemize}}
\newenvironment{myenumerate}{\begin{enumerate}}{\end{enumerate}}

\IGNORE{
\newcounter{i}
\newenvironment{mylist}[1]{\begin{list}{#1}{\usecounter{i}%
   \setlength{\leftmargin}{3ex}}\setlength{\labelwidth}{1.5ex}%
   \setlength{\labelsep}{1.5ex}}{\end{list}}
\newenvironment{myelist}[1]{\begin{list}{#1}{\usecounter{i}%
   \setlength{\leftmargin}{5ex}}\setlength{\labelwidth}{3.75ex}%
   \setlength{\labelsep}{1.25ex}}{\end{list}}

\renewenvironment{myenumerate}{\begin{myelist}{\arabic{i}.}}{\end{myelist}}
\renewcommand{\paragraph}[1]{{\medskip\par\noindent\bf #1}}
\renewenvironment{pf}{\begin{proof}}{\hspace*{\fill}\end{proof}}
}

\begin{document}

\maketitle

\begin{abstract}
  We reexamine fundamental problems from computational geometry in the
  word RAM model, where input coordinates are integers that fit in a
  machine word. We develop a new algorithm for offline point location,
  a two-dimensional analog of sorting where one needs to order points
  with respect to segments. This result implies, for example, that the
convex hull of $n$ points in three dimensions can be constructed in
  (randomized) time $n\cdot 2^{O(\sqrt{\lg\lg n})}$.  Similar bounds
  hold for numerous other geometric problems, such as
planar Voronoi diagrams,
planar off-line nearest neighbor search,
  line segment intersection, and triangulation of non-simple
  polygons.

  In FOCS'06, we developed a data structure for \emph{online} point
  location, which implied a bound of $O(n \frac{\lg n}{\lg\lg n})$ for
three-dimensional convex hulls and the other problems. Our current bounds are
  dramatically better, and a convincing improvement over the classic
  $O(n\lg n)$ algorithms. As in the field of integer sorting, the main
  challenge is to find ways to manipulate information, while avoiding
  the online problem (in that case, predecessor search).

\IGNORE{
}
\end{abstract}


\section{Introduction}

\subsection{Sorting in Two Dimensions}

Consider the following toy problem (in fact, a special case of offline
point location), which we call \emph{the slab problem}. We are given a
vertical slab in the plane, $m$ nonintersecting segments cutting
across the slab, and $n$ points in the slab. The goal is to identify
the segment immediately below each of the $n$ points. In other words,
we would like to sort the points ``relative to'' the segments.

This is an appealing and natural generalization to two dimensions of
the one-dimensional notion of sorting. It captures both an intuitive
notion of ordering, and the non-orthogonal flavor so common in
computational geometry.
Indeed,
as described below, an impressive collection of fundamental problems
in computational geometry are known to be reducible to this simple
problem, so there is a formal sense in which the slab problem is as
central in computational geometry as sorting is in the one-dimensional
world. 

Classically, the slab problem is solved by binary searching among
segments for each input point, for a cost of $O(\lg m)$ per point.
This is optimal when one searches by binary decisions or assumes the
input has infinite precision, as on a real RAM. However, a more reasonable
assumption is that input has finite precision. We will assume, in
particular, that all coordinates come from some universe $[2^w]=\{ 0, \dots,
2^w-1 \}$, and that we are working on a word RAM with $w$-bit words
(i.e.~one coordinate fits in one word). See Section \ref{sec:why-ints}
for a discussion of these assumptions.

Until recently, successful use of the word RAM in computational
geometry was limited to a restricted class of problems, especially
problems involving orthogonal objects. However, in FOCS'06, we
proposed improved data structures for the
\emph{online} slab problem, a problem of a fundamentally nonorthogonal
nature~\cite{ChaPat}. 
The running time was asymptotically $\min \left\{ \frac{\lg
m}{\lg\lg m}, \sqrt{\frac{w}{\lg w}} \right\}$ per point. This
represents a marginal improvement over $O(\lg m)$ for \emph{any}
universe, and a roughly quadratic improvement for small (polynomial)
universes.

In the current paper, we describe an algorithm for the (offline) slab
problem running in time $n \cdot 2^{O(\sqrt{\lg\lg m})} + O(m)$. Note
that this bound does not depend on the universe (aside from assuming a
coordinate fits in a word), and is deterministic. The bound is a
dramatic improvement over our old bounds---note, for example, that
the new bound grows more slowly than $n \lg^\eps m + m$ for any constant
$\eps>0$. In addition, the new bound represents a much more convincing
improvement over the standard $O(n\lg m)$ bound, demonstrating the
power granted by bounded precision.

The relation between our current algorithm and our results from
\cite{ChaPat} is best understood by a parallel to
integer sorting. There, the online problem (predecessor search) is
known to require comparatively large running times (e.g.~in terms
of $n$ alone, an $\Omega(\sqrt{\frac{\lg n}{\lg\lg n}})$ lower bound per
point is known~\cite{BeaFic}).  Yet, one 
can find
ways of manipulating information in the offline problem, such that the
bottleneck of using the online problem is avoided (e.g.~we can
sort in $O(n\sqrt{\lg\lg n})$ expected time~\cite{HanThoFOCS02}). It should 
be
understood that the purpose of this work is not to study ``bit
tricks'' in the word RAM model, but to study how information about
points and lines can be decomposed in algorithmically useful ways.

\subsection{Applications}

From \cite{ChaPat} it follows that improved bounds
for the slab problem lead to improved upper bounds for many
fundamental problems in computational geometry 
\cite{BerBOOK,EdeBOOK,MulBOOK,OroBOOK,PreSha}. We list some here. As
before, the bounds do not depend on the universe for the coordinates.
All the reductions below, except the last, are randomized.

\begin{myenumerate}
\item We can compute the {\em convex hull\/} of $n$ points in three
  dimensions in expected time $n\cdot 2^{O(\sqrt{\lg\lg n})}$.  If the
  hull has $H$ vertices, the bound can be reduced to $n\cdot
  2^{O(\sqrt{\lg\lg H})}$.

\item We can compute the {\em Voronoi diagram\/} and the {\em Delaunay
   triangulation\/} of $n$ points in the plane in expected time
   $n\cdot 2^{O(\sqrt{\lg\lg n})}$.  As a consequence, we can also
   compute the {\em Euclidean minimum spanning tree\/} or solve the
   {\em largest empty circle\/} problem within the same time bound.

\item Given $n$ red points and $n$ blue points in the plane, we can
  compute the red point nearest to each blue point in expected time
  $n\cdot 2^{O(\sqrt{\lg\lg n})}$.

\item We can compute all $K$ intersections of $n$ line segments in the
  plane in expected time $n\cdot 2^{O(\sqrt{\lg\lg n})} + O(K)$. We
  can also construct the {\em trapezoidal decomposition\/} of the line
  segments within the same time bound.

\item We can triangulate a polygon with holes (or an environment with
  multiple disjoint polygons) with $n$ vertices total in 
  time $n\cdot 2^{O(\sqrt{\lg\lg n})}$.
\end{myenumerate}


Problems like convex hulls and Voronoi diagrams date back to the dawn
of computational geometry, and for these problems standard $O(n\lg n)$
bounds have long been regarded as ``optimal.''  The previous paper
\cite{ChaPat} has demonstrated that on the word RAM, $O(n\lg n)$
can be beaten slightly, by $O(n\frac{\lg n}{\lg\lg n})$ bounds.  The current
paper shows that $O(n\lg n)$ can be improved significantly.

\paragraph{Planar point location.}
The slab problem is actually a special case of the {\em offline planar
point location\/} problem.  Here, the input consists of a connected
polygonal subdivision defined by a set of $m$ segments in the plane,
where segments are only allowed to touch at endpoints. Given $n$ query
points (offline), the goal is to identify the polygon (face) which
contains each point.


In fact, reductions discussed above are to the offline planar point
location problem. However, the general case of point location turns
out to be reducible to the special case of the slab problem. In
\cite{ChaPat}, we considered three different ways to
achieve this reduction. These three approaches hold both in the
offline and online cases. They all generate a multiplicative penalty
of at most $O(\lg\lg m)$ per point, which is absorbed by our time
bound, but differ in the cost per segment:
\begin{myenumerate}
\item[(i)] random sampling gives a \emph{randomized} $(n+m)\cdot
  2^{O(\sqrt{\lg\lg m})}$ running time.

\item[(ii)] persistence and exponential trees give a deterministic
  $n\cdot 2^{O(\sqrt{\lg\lg m})} + O(\textrm{sort}(m))$ bound, where
  $\textrm{sort}(m)$ denotes the cost of sorting $m$ integers.

\item[(iii)] planar separators are the most complicated (and least
  practical) strategy but give the best deterministic running time of
  $n \cdot 2^{O(\sqrt{\lg\lg m})} + O(m)$.
\end{myenumerate}

\paragraph{Higher dimensions.}
We can also solve the analog of the slab problem in any constant
dimension. Instead of a slab, we are given a vertical prism, and
instead of segments, we are given hyperplanes cutting across the prism
where no two hyperplanes intersect inside the prism. The running time of
our solution is $n\cdot 2^{O(\sqrt{\lg\lg m})} \lg^{1+o(1)} w + O(m)$.
Although the bound now has an extra $\lg^{1+o(1)}w$ factor, this
factor is relatively small.  This result has a few applications as
well, for example, to offline exact nearest neighbor search in higher
dimensions and curve-segment intersection in two dimensions (see
\cite{ChaPat}).

\paragraph{Recent work.}
Subsequent to our conference publication, Buchin and
Mulzer~\cite{BucMul} announced a better result for planar Voronoi
diagrams. They show that constructing Voronoi diagrams is equivalent
to sorting, in the Word RAM augmented with one non-standard
operation. In the standard Word RAM, they obtain an unconditional
running time of $O(n \sqrt{\lg\lg n})$ by adapting the best known
sorting algorithm of~\cite{HanThoFOCS02}. 

This improves the running time of all problems mentioned in item
2.~above. However, this algorithm exploits special properties about
Voronoi diagrams and nearest neighbor graphs, and does not imply
improved results for the other problems considered here, such as 3-d
convex hulls and offline point location.

\subsection{Computational Geometry on a Grid}  \label{sec:why-ints}

It is superfluous to state that bounded precision is a fact of
life. Input data are given with finite precision and computers
represent it internally in a bounded number of bits. Low-dimensional
computational geometry has typically seen the negative consequences of
this state of affairs. Algorithms are designed in idealized models of
real arithmetic, and practitioners struggle to keep the algorithms
working with imperfect precision.

Yet, there is also a good side to bounded precision, and our result
shows it can be used to achieve significantly better
algorithms. Again, we emphasize that our improved bound is independent
of whatever the bound on the precision happens to be. We just require
that coordinates can be manipulated in constant time, which has always
been a standard assumption.

The philosophical question that we wish to address briefly is whether
these benefits of bounded precision should be explored in theory. We
believe firmly they should, examining the question both with an eye to
practice and to theory.

\paragraph{Practice.}
From a theoretical perspective, the classic solutions to online point
location using linear space and logarithmic query time would seem
attractive. However, as pointed out in a survey on the topic
\cite{SnoSURV}, the most efficient and popular practical
implementations do not use them. Instead, they use grid-pruning
heuristics, not unlike some of our ideas. Thus, we can hope to gain
theoretical understanding for what is already known to be effective in
the real world---a standard goal for theory.

Turning this around, we can hope that theoretical improvements will
suggest new ideas with an impact in practice. As presented, our
results are theoretical because of large hidden constants, in both the
slab problem and subsequent reductions. However, since the improvement
over $O(n\lg n)$ is now quite significant, we find it plausible that
some of the techniques developed here can provide inspiration for
useful practical ``tricks.'' A key step would be to circumvent the
reductions and apply our techniques directly to target problems.

\paragraph{Theory.}
Even at theory's end of computational geometry, the assumption of
bounded precision has been used fairly often. Unfortunately, this body
of work (see the bibliographies of \cite{ChaPat})
has been plagued by close ties to one-dimensional
problems. For example, two-dimensional convex hulls can be found in
linear time once points are sorted by $x$-coordinate. More typically,
the problems considered involved orthogonal objects, and such
orthogonal problems can more easily be decomposed into one-dimensional
problems.

Our recent data structures for point location \cite{ChaPat} broke this barrier, by presenting an improvement for a
fundamentally nonorthogonal problem. The current paper tries to
demonstrate that there are deeper questions to be explored in this
direction of research. In our algorithm, we are forced to consider
questions of decomposability and compressibility of information about
two-dimensional objects, which seem fundamentally different from
questions in one dimension. We feel this should have a theoretical
appeal in itself.

For example, consider two standard tools in sorting. One is radix
sort, which gives linear-time sorting in polynomial universes.
Another is hashing, which is used in virtually all advanced RAM
sorting algorithms, including
\cite{AndFOCS96,AndJCSS,HanINFCOMP,HanSTOC02,HanThoFOCS02,KirRei,ThoSODA97}.
In two dimensions, neither of these tools seems
relevant. 
It is interesting that even without such basic
tools, we can still obtain a rather efficient, determinstic algorithm
(even outperforming some of the older RAM sorting results).


\IGNORE{
%
%
\paragraph{What kind of bounded precision?}
The final question that we wish to touch on is whether an integer
universe is the right model for bounded precision. In certain cases,
the input is on an integer grid by definition (e.g.~objects are on a
computer screen). One might worry, however, about the input being a
floating point number. We believe that in most cases this is an
artifact of representation, and numbers should be treated as integers
after appropriate scaling. One reason is to note that the
``floating-point plane'' is simply a union of bounded integer grids
(the size depending on the number of bits of the mantissa), at
different scale factors around the origin. Since our problems are
translation-independent, there is no reason the origin should be
special, and having more detail around the origin is not particularly
meaningful. Another reason is that certain aspects of the problems are
not well-defined when inputs are floating point numbers. For example,
the slope of a line between two points of very different exponents is
not representable by floating point numbers anywhere close to the
original precision.
}

\subsection{Overview}

The remainder of this paper is organized as follows. In Section
\ref{sec:sqrt-lg}, we describe a simple algorithm for the slab
problem, running in time $O(n \sqrt{\lg m} + m)$. This demonstrates
the basic divide-and-conquer strategy behind our solution. In Section
\ref{sec:better}, we implement this strategy
much more carefully to obtain an interesting recurrence 
that ultimately leads to the stated time bound of $n \cdot
2^{O(\sqrt{\lg\lg m})} +O(m)$. The challenges faced by this improvement are
similar to issues in integer sorting, and indeed we borrow (and build 
upon) some tools from that field.

Unfortunately, the implementation of Section \ref{sec:better} requires
a nonstandard word operation. In Section \ref{sec:stdops}, we describe
how to implement the algorithm on a standard word RAM,
using only addition, multiplication, bitwise-logical 
operations, and shifts.  Interestingly, the new implementation requires 
some new geometric observations that affect the
design of the recursion 
itself.

\section{An Initial Algorithm for the Slab Problem} \label{sec:sqrt-lg}

\subsection{The Basic Recursive Strategy}

We begin with a recursive strategy based on a simple observation,
taken from~\cite{ChaPat}. (Later in Section \ref{sec:stdops}, we will
replace this with a more complicated recursive structure.)
In the following, the notation $\prec$ refers to the belowness
relation.

\newcommand{\sss}{\tilde{s}}

\begin{observation} \label{obs:basic}
Fix $b$ and $h$. Let $S$ be a set of $m$ sorted disjoint segments,
where all left endpoints lie on an interval $I_L$ of length
$2^{\ell_L}$ on a vertical line, and all right endpoints lie on an
interval $I_R$ of length $2^{\ell_R}$ on another vertical line.  In
$O(b)$ time, we can find $O(b)$ segments $s_0,s_1,\ldots\in S$ in
sorted order, which include the lowest segment of $S$, such that:
\begin{enumerate}
\item[\em (1)] for each $i$, at least one of the following holds:
  \begin{enumerate}
  \item[\em (1a)] there are at most $m/b$ segments of $S$ between $s_i$
    and $s_{i+1}$.
  \item[\em (1b)] the left endpoints of $s_i$ and $s_{i+1}$ lie on a
    subinterval of length $2^{\ell_L-h}$.
  \item[\em (1c)] the right endpoints of $s_i$ and $s_{i+1}$ lie on a
    subinterval of length $2^{\ell_R-h}$.
  \end{enumerate}

\item[\em (2)] there exist segments $\sss_0,\sss_2,\ldots$ cutting
  across the slab, satisfying all of the following:
  \begin{enumerate}
    \item[\em (2a)] $s_0 \prec \sss_0 \prec s_2 \prec \sss_2 \prec
      \cdots$.
    \item[\em (2b)] distances between the left endpoints of the
      $\sss_i$'s are all multiples of $2^{\ell_L-h}$.
    \item[\em (2c)] distances between right endpoints are all
      multiples of $2^{\ell_R-h}$.
  \end{enumerate}
\end{enumerate}
\end{observation}

\begin{figure}
\begin{center}
\setlength{\unitlength}{1.6in}
\input{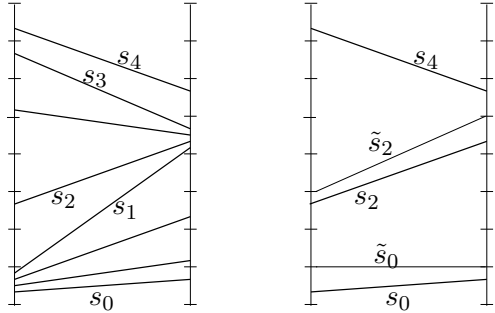}
\end{center}\vspace{-2.5ex}
\caption{Proof of Observation \protect\ref{obs:basic}: an example
(with $\ell_L=\ell_R$).
The left diagram shows the segments in $B$.}
\label{fig:slab}
\end{figure}

\newcommand{\slab}{\mbox{\sc Slab}}
\newcommand{\rnd}{\mbox{\sc round}}
\newcommand{\ans}{\mbox{\sc ans}}
\newcommand{\ansnull}{\mbox{\sc null}}

\begin{figure*}
\center
\fbox{\hspace{-.5cm}
\begin{minipage}{7cm}
  \begin{algm} $\slab(Q,S)$:\\[2pt]
    0.\' if $m=0$, set all answers to $\ansnull$ and return \\
    1.\' let $s_0, s_1, \ldots$ be the $O(b)$ segments from
      Observation \ref{obs:basic} \\
    2.\' let $\varphi$ be the projective transform mapping $I_L$ to
      $\{0\}\times [0,2^h]$ and $I_R$ to $\{2^h\}\times [0,2^h]$. \\
      \> Compute $\rnd(\varphi(Q))$ and
           $\varphi(\sss_0),\varphi(\sss_2),\ldots$ \\
    3.\' $\slab_0(\rnd(\varphi(Q)),\:
      \{ \varphi(\sss_0),\varphi(\sss_2),\ldots \})$\\
    4.\' for each $q\in Q$ with
      $\ans[\rnd(\varphi(q))]=\varphi(\sss_i)$ do\\ \> set $\ans[q]=$
      the segment from $\{s_{i-4}\TO s_{i+7}\}$ immediately below
      $q$\\
    5.\' for each $s_i$ do\\
      \> $\slab(\{q\in Q\mid \ans[q]=s_i\},\:
      \{s\in S\mid s_i\prec s\prec s_{i+1}\})$
  \end{algm}
\end{minipage}
\hspace{.5cm}}
\caption{A recursive algorithm for the slab problem. Parameters $b$
  and $h$ are fixed in the analysis; $\rnd(\cdot)$ maps a point to
  its nearest integral point.}
\label{fig:slab-code}
\end{figure*}

\begin{pf}
Let $B$ contain every $\down{m/b}$-th segment of $S$, starting with
the lowest segments $s_0$. Impose a grid over $I_L$ consisting of
$2^h$ subintervals of length $2^{\ell_L-h}$, and a grid over $I_R$
consisting of $2^h$ subintervals of length $2^{\ell_R-h}$. We define
$s_{i+1}$ inductively based on $s_i$, until the highest segment is
reached. We let $s_{i+1}$ be the highest segment of $B$ such that
either the left or the right endpoints of $s_i$ and $s_{i+1}$ are in
the same grid subinterval. This will satisfy (1b) or (1c). If no such
segment above $s_i$ exists, we simply let $s_{i+1}$ be the successor
of $s_i$, satisfying (1a).  (See Figure~\ref{fig:slab}.)

Let $\sss_i$ be obtained from $s_i$ by rounding each endpoint to the
grid point immediately above (ensuring (2b) and (2c)). By construction
of the $s_i$'s, both the left and right endpoints of $s_i$ and $s_{i+2}$
are in different grid subintervals. Thus, $\sss_i \prec s_{i+2}$,
ensuring (2a).
\end{pf}

\bigskip
The above observation naturally suggests a recursive algorithm.
In the pseudocode in
Figure \ref{fig:slab-code}, the input is a set $Q$
of $n$ points and a sorted set $S$ of $m$ disjoint segments, where the
left and right endpoints lie on intervals $I_L$ and $I_R$ of length
$2^{\ell_L}$ and $2^{\ell_R}$ respectively. At the end, $\ans[q]$
stores the segment from $S$ immediately below $q$ for each $q\in Q$.
A special $\ansnull$ value for $\ans[q]$ signifies that $q$ is below all
segments. We assume a (less efficient) procedure $\slab_0(Q,S)$, with
the same semantics as $\slab(Q,S)$, which is used as a bottom case of
the recursion. The choice of $\slab_0()$ is a crucial component of the
analysis.

\medskip

We first explain why the pseudocode works.  In step~2, an explicit
formula for the transform $\varphi$ has already been given
in~\cite{ChaPat}; this mapping preserves the belowness relation.
According to property (2) in Observation \ref{obs:basic}, we know that
the transformed segments $\varphi(\sss_0), \varphi(\sss_2), \ldots$
all have $h$-bit integer coordinates from $[2^h]$. After rounding, the
$n$ points $\varphi(Q)$ will lie in the same universe.

Any unit square can intersect at most two of the $\varphi(\sss_i)$'s,
since these segments have vertical separation at least one and thus
horizontal separation at least one (as slopes are between $-1$ and
$1$).  If $\varphi(\sss_i) \prec \rnd(\varphi(q)) \prec
\varphi(\sss_{i+2})$, then we must have $\varphi(\sss_{i-4}) \prec
\varphi(q) \prec \varphi(\sss_{i+6})$, implying that $s_{i-4} \prec
\sss_{i-4} \prec q \prec \sss_{i+6} \prec s_{i+8}$.  Thus, at step~4,
$\ans[q]$ contains the segment from $s_0,s_1\ldots$ immediately below
$q$.  Once this is determined for every point $q\in Q$, we can
recursively solve the subproblem for the subset of points and segments
strictly between $s_i$ and $s_{i+1}$ for each $i$, as is done at
step~5.  An answer $\ans[q] = \ansnull$ from the $i$-th subproblem is
interpreted as $\ans[q] = s_i$.

Let $\ell = (\ell_L + \ell_R)/2\ (\ell\le w)$. Denote by $T(n,m,\ell)$ the 
running
time of $\slab()$, and $T_0(n,b',h)$ the running time of the call to
$\slab_0()$ in step~3.  Steps~1, 2 and~4 can be implemented naively in
$O(n+m)$ time.  We have the recurrence:
\begin{equation} \label{eq:basic-rec}
T(n,m,\ell)\ =\ T_0(n,b',h) \:+\: O(n+m) \:+\: 
\sum_{i=0}^{b'} T(n_i,m_i,\ell_i),
\end{equation}
where $b'=O(b)$, $\sum_i n_i=n$, $\sum_i m_i = m-b'$. Furthermore,
according to property (1) in Observation~\ref{obs:basic}, for each $i$
we either have $m_i\le \frac{m}{b}$ or $\ell_i \le \ell-\frac{h}{2}$.
This implies that the depth of the recursion is $O(\log_b m +
\frac{\ell}{h})$.

\subsection{An $O(n\sqrt{\lg m} + m)$ Algorithm}

In the previous paper~\cite{ChaPat}, we have noticed that for $b'
h\approx w$, $\slab_0()$ can be implemented in $T_0(n,b',h)=O(n)$ time
by packing $b'$ segments from an $h$-bit universe into a word.
By setting $b\approx \log^\eps m$ and $h\approx w/\log^\eps m$,
this leads to an $O((n + m) \frac{\lg m}{\lg\lg m})$ algorithm.

Instead of packing multiple segments in a word, our new idea is to
pack {\em multiple points\/} in a word.  To understand why this helps,
remember that the canonical implementation for $\slab_0()$ runs in
time $O(n\lg m)$ by choosing the middle segment and recursing on
points above and below this segment. By packing $t$ segments in a
word, we can hope to reduce this time to $O(n\log_t m)$. However, by
packing $t$ points in a word, we can potentially reduce this to
$O(\frac{n}{t} \lg m)$, a much bigger gain. (One can also think of
packing both points and segments, for a running time of $O(\frac{n}{t}
\log_t m)$. Since we will ultimately obtain a much faster algorithm,
we ignore this slight improvement.)

To implement this idea, step 2 will pack $\rnd(\varphi(Q))$ with $O(w
/ h)$ points per word. Each point is alotted $O(h)$ bits for the
coordinates, plus $\lg b = O(h)$ bits for the answer
$\ans[\rnd(\varphi(q))]$ which $\slab_0()$ must output. This packing
can be done in $O(n)$ time, adding one point at a time.

Working on packed points, $\slab_0()$ has the potential of running
faster, as evidenced by the following lemma. For now, we do not
concern ourselves with the implementation on a word RAM, and assume
nonstandard operations (an operation takes two words as input, and
outputs one word).

\begin{lemma}   \label{lem:basic-zero}
If $\lg b \le h\le w$,
$\slab_0()$ can be impemented on a RAM with nonstandard operations
with a running time of $T_0(n,b,h) = O(n \frac{h}{w} \lg
b + b)$.
\end{lemma}

\begin{pf}
Given a segment and a number of points packed in a word, we can
postulate two operations which output the points above (respectively
below) the segment, packed consecutively in a word. Choosing a
segment, we can partition the points into points above and below the
segment in $O(\lceil n \frac{h}{w} \rceil)$ time. In the same
asymptotic time, we can make both output sets be packed with $\lfloor
\frac{w}{h} \rfloor$ points per word (merging consecutive words which
are less than full).

We now implement the canonical algorithm: partition points according
to the middle segment and recurse. As long as we are working with $\ge
\frac{w}{h}$ points, the cost is $O(\frac{h}{w})$ per point for each
segment, and each point is considered $O(\lg b)$ times. If we are
dealing with less than $\frac{w}{h}$ points, the cost is $O(1)$, and
that can be charged to the segment considered. Thus, the total time
after packing is $O(n\frac{h}{w}\lg b + b)$.

The last important issue is the representation of the output. By the
above, we obtain the sets of points which lie between two consecutive
segments. We can then trivially fill in the answer for every point in
the $\lg b$ bits alotted for that.  However, we want an array of
answers for the points in the original order. To do that, we trace the
algorithm from above backwards in time. We use an operation which is
the inverse of splitting a word into points above and below a segment.
\end{pf}

\bigskip

Plugging the lemma into (\ref{eq:basic-rec}), we get $T(n, m,\ell) =
O( n \frac{h}{w}\lg b + n + m ) \cdot O(\log_b m +
\frac{\ell}{h})$.  Setting $\lg b = \sqrt{\lg m}$ and $h= w /
\sqrt{\lg m}$, we obtain $T(n,m,w)=O((n+m) \sqrt{\lg m})$.  This can
be improved to $O(m + n\sqrt{\lg m})$ by the standard trick of
considering only one in $O(\sqrt{\lg m})$ consecutive segments. For
every point, we finish off by binary searching among $O(\sqrt{\lg m})$
segments, for a negligible additional time of $O(n\lg\lg m)$.

\section{An $n \cdot 2^{O(\sqrt{\lg\lg m})} + O(m)$ Algorithm}  
\label{sec:better}

\subsection{Preliminaries}

To improve on the $O(m+n\sqrt{\lg m})$ bound, we {\em bootstrap\/}: we use
an improved algorithm for $\slab()$ as $\slab_0()$, obtaining an even
better bound for $\slab()$. To enable such improvements, we can no
longer afford the $O(n)$ term in the recurrence (\ref{eq:basic-rec}).
Rather, a call to $\slab()$ is passed $Q$ in word-packed form, and we
want to implement the steps between recursive calls in {\em
sub\/}linear time (close to the number of words needed to represent
$Q$, not to $n=|Q|$).

This task will require further ideas and more sophisticated
word-packing tricks. To understand the complication, let us contrast
implementing steps~2 and~5 of $\slab()$ in sublinear time. Computing
$\rnd(\varphi(Q))$ in Step 2 is solved by applying a function in
parallel to a word-packed vector of points. This is certainly
possible, at least using nonstandard word operations. However, step 5
needs to group elements of $Q$ into subsets (i.e.~sort $Q$ according
to $\ans[q]$). This is a deeper information-theoretic limitation, and
it is rather unlikely that it can always be done in time linear in the
number of words needed to store $Q$. The problem has connections to
applying permutations in external memory, a well-studied problem which
is believed to obey similar limitations \cite{AggVit}.

\newcommand{\Split}{\mbox{\sc Split}}
\newcommand{\Unsplit}{\mbox{\sc Unsplit}}
\newcommand{\Label}{\mbox{\sc label}}

\medskip

To implement step 5 (and also step 4), we will use a subroutine
$\Split(Q, \Label)$. This receives a set $Q$ of $\ell$-bit elements,
packed in $O(n\frac{\ell}{w})$ words. Each elements $q\in Q$ has a
$(\lg m)$-bit label $\Label[q]$ with $\lg m\le\ell$.  The labels are stored
in the same $O(n\frac{\ell}{w})$ words.  We can think of each word
as consisting of two portions, the first containing $O(\frac{w}{\ell})$
elements and the second containing the corresponding 
$O(\frac{w}{\ell})$ labels.
The output of $\Split(Q, \Label)$ is a collection
of sublists, so that all elements of $Q$ with the same label are put
in the same sublist (in arbitrary order).

In addition, we will need $\Split()$ to be reversible. Suppose the
labels in the sublists have been modified. We need a subroutine
$\Unsplit(Q)$, which outputs $Q$ in the original order before
$\Split()$, but with the modified labels attached.

The following lemma states the time bound we will use for these two
operations. The implementation of $\Split()$ is taken from a paper by
Han~\cite{HanINFCOMP} and has been also used as a subroutine in
several integer sorting algorithms \cite{HanSTOC02,HanThoFOCS02}.  As
far as we know, the observation that $\Unsplit()$ is possible in the
same time bound has not been stated explicitly before.

\begin{lemma} \label{lem:split}
Assume $\Label[q] \in [m]$ for all $q\in Q$, and let $M$ be a
parameter. If $\frac{w}{\ell}\lg m \le \frac{1}{2} \lg M$ and $\lg m
\le \ell \le w$, both $\Split()$ and $\Unsplit()$ require time
$O(n\frac{\ell}{w}\lg \frac{w}{\ell} + M)$.
\end{lemma}

\begin{pf}
Let $g = \frac{w}{\ell}$. Each word contains $g$ elements, with $g\lg
m$ bits of labels. Put words with the same label pattern in the same
bucket. This can be done in $O(n/g + \sqrt{M})$ time, since the number
of different label patterns is at most $2^{g\lg m}\le \sqrt{M}$.  For
each bucket, we form groups of $g$ words and {\em transpose\/} each
group to get $g$ new words, where the $i$-th element of the $j$-th new
word is the $j$-th element of the $i$-th old word.  Transposition can
be implemented in $O(\lg g)$ standard word operations~\cite{ThoSODA97}.
Elements in each new word now have identical labels.  We can put these
words in the correct sublists, in $O(n/g + m)$ time.  There are at
most $g$ leftover elements per bucket, for a total of
$O(\sqrt{M}g)=o(M)$; we can put them in the correct sublists in $o(M)$
time.  The total time is therefore $O((n/g)\lg g + M)$.

To support unsplitting, we remember information about the splitting
process.  Namely, whenever we transpose $g$ words, we create a record
pointing to the $g$ old words and the $g$ new words.  To unsplit, we
examine each record created
and transpose its $g$ new words again to get back
the $g$ old words (with labels now modified).  We can also update
the leftover elements by creating $o(M)$ additional pointers.
\end{pf}

\bigskip

A particularly easy application of this machinery is to implement the
algorithm of Section \ref{sec:sqrt-lg} with standard operations (with
a minor $\lg\lg m$ slowdown). This result is not interesting by
itself, but it will be used later as the base case of our
bootstrapping strategy.

\begin{corollary}  \label{cor:zero}
If $\frac{w}{h}\lg b \le \frac{1}{2} \lg M$ and $\lg b\le h\le w$,
the algorithm for $\slab_0()$ from Lemma \ref{lem:basic-zero} can be
implemented on a word RAM with standard operations in time 
$T_0(n,b,h)=O(n \frac{h}{w} \lg b\,\lg \frac{w}{h} + bM)$.
\end{corollary}

\begin{pf}
The nonstandard operations used before were splitting and unsplitting
a set of points packed in a word, depending on sidedness with respect
to a segment. It is not hard to compute sidedness of all points from a
word in parallel using standard operations: we apply the linear map
defining the support of the segment to all points (which is a parallel
multiplication and addition), and keep the sign bits of each
result. The sign bits define 1-bit labels for the points, and we can
apply $\Split()$ and $\Unsplit()$ for these.
%
\end{pf}

\bigskip

Since the algorithm is used with $b=\sqrt{\lg m}$ and $h= w/\sqrt{\lg m}$, 
we
incur a slowdown of $O(\lg\frac{w}{h})=O(\lg\lg m)$ per point compared to 
the implementation 
with nonstandard operations. By setting $M=m^2$, 
the algorithm of the previous section would then 
run in time $O(n\sqrt{\lg m} \lg\lg m + m^3)$ if implemented with
standard operations.  (The dependence of the second term on $m$ can be 
lowered as well.)

\subsection{The Improved Algorithm}

\newcommand{\elll}{{\widetilde{\ell}}}
\newcommand{\mm}{\widetilde{m}}

Our fastest algorithm follows the same pseudocode of Figure
\ref{fig:slab-code}, but with a more careful implementation of the
individual steps. Let $\elll$ be the number of bits per point and
$\mm$ the original number of segments in the root call to $\slab()$.
We have $\lg\mm\le\elll\le w$. In a recursive call to $\slab()$, the
input consists of some $n$ points and $m\le\mm$ segments, all with
coordinates from $[2^\ell]$, where $\ell\le \elll$. Points will be
packed in $O(\elll)$ bits each, so the entire set $Q$ occupies
$O(n\frac{\elll}{w})$ words. At the end, the output $\ans[q]$ is
encoded as a label with $\lg\mm$ bits, stored within each point $q\in
Q$, with the order of the points unchanged in the list~$Q$. Note that
one could think of repacking more points per word as $\ell$ and
$m$ decrease, but this will not yield an asymptotic advantage, so
we avoid the complication (on the other hand, repacking before the
call to $\slab_0()$ is essential).

In step~2, we can compute $\rnd(\varphi(Q))$ in time linear in the
number of words $O(n\frac{\elll}{w})$, by using a nonstandard word
operation that applies the projective transform (and rounding) to
multiple points packed in a word. Unfortunately, it does not appear
possible to implement this efficiently using standard operations. We
will deal with this issue in Section \ref{sec:stdops}, by changing the
algorithm for $\slab()$ so that we only require affine
transformations, not projective transformations.

Before the call to $\slab_0()$ in step~3, we need to condense the
packing of the points $\rnd(\varphi(Q))$ to take up $O(n\frac{h}{w})$
words. Previously, we had $O(\frac{w}{\elll})$ points per word, but
after step~2, only $O(h)$ bits of each point were nonzero.  We will
stipulate that points always occupy an number of bits which is a power
of 2.  This does not affect the asymptotic running time.  Given this
property, we obtain a word of $\rnd(\varphi(Q))$ by condensing $\elll
/ h$ words.  This operation requires shifting each old word, and {\sc
or}ing it into the new word.

Note that the order of $\rnd(\varphi(Q))$ is different from the order
of $Q$, but this is irrelevant, because we can also reverse the
condensing easily. We simply mask the bits corresponding to old word,
and shift them back. Thus, we obtain the labels generated by
$\slab_0()$ in the original order of $Q$. Both condensing and its
inverse take $O(n \frac{\elll}{w})$ time.

For the remainder of the steps, we need to $\Split()$ and $\Unsplit()$.
For that, we fix a parameter $M$ satisfying $\frac{w}{\elll} \lg \mm
\le \frac{1}{2} \lg M$. In step~4, we first split the list
$\rnd(\varphi(Q))$ into sublists with the same $\ans$ labels. For each
sublist, we can perform the constant number of comparisons per point
required in step~4, and then record the new $\ans$ labels in the list,
in time linear in the number of words $O(n\frac{\elll}{w})$. It is
standard to implement this in the word RAM by parallel multiplications
(see the proof of Lemma \ref{lem:split}). To complete step 4, we
$\Unsplit()$ to get back the entire list $\rnd(\varphi(Q))$, and then
copy the $\ans$ labels to the original list $Q$ in
$O(n\frac{\elll}{w})$ time. Since both lists are in the same order,
this can be done by masking labels and {\sc or}ing them in.

To perform step~5, we again split $Q$ into sublists with the same
$\ans$ labels.  After the recursive calls, we unsplit to get $Q$ back
in the original order, with the new $\ans$ labels.

\subsection{Analysis}

For $\frac{w}{\elll}\lg\mm\le\frac{1}{2}\lg M$ and
$\lg\mm\le\elll\le w$,
the recurrence (\ref{eq:basic-rec}) now becomes:

\begin{equation}  \label{eq:new-rec}
T(n,m,\ell)\ =\ T_0(n,b',h) 
\:+\: O\left(n\frac{\elll}{w}\lg\frac{w}{\elll} + M\right)
\:+\: \sum_{i=0}^{b'} T(n_i,m_i,\ell_i),
\end{equation}
%
where $b'=O(b)$, $\sum_i n_i=n$, $\sum_i m_i = m-b'$, and for each
$i$, we either have $m_i\le \frac{m}{b}$ or $\ell_i \le \ell -
\frac{h}{2}$. As before, the depth of the recursion is bounded by
$O(\log_b \mm + \frac{\elll}{h})$.

Assume that for $\frac{w}{h} \lg b \le \frac{1}{2}\lg M$ and $\lg b
\le h \le w$, an algorithm with running time
\[ T_0(n,b,h)\ \le\ c_k
   \left(n\frac{h}{w} \lg^{1/k} b\:\lg\left(\frac{w}{h}\lg b\right) 
   \:+\: bM\right)
\]
is available to begin with. This is true for $k=1$ with $c_1=O(1)$
by Corollary \ref{cor:zero}.

Then the recurrence (\ref{eq:new-rec}) yields:

\[ T(n,\mm,\elll)\ =\ O(c_k) \cdot \left(
   \left[ n\frac{h}{w} \lg^{1/k} b\:\lg\left(\frac{w}{h} \lg b\right) 
          \:+\: n \frac{\elll}{w} \lg \frac{w}{\elll} \right]
   \cdot \left( \log_b \mm + \frac{\elll}{h} \right) \:+\: mM\right).
\]
%

Set $\lg b = \lg^{k/(k+1)} \mm$ and $h = \elll \mbox{\big /}
\lg^{1/(k+1)} \mm$.  Notice that indeed $\frac{w}{h} \lg b =
\frac{w}{\elll} \lg\mm \le \frac{1}{2} \lg M$ and $\lg b \le h \le w$.
Thus, we obtain an algorithm with running time:
\[ T(n,\mm,\elll)\ \le\ c_{k+1} \left(
   n \frac{\elll}{w}\log^{1/(k+1)} \mm\:\lg\left( \frac{w}{\elll} \lg\mm \right)
   \:+\: \mm M\right)
\]
for some $c_{k+1}=O(1) \cdot c_k$.

Iterating this process $k$ times, we get:
\[ T(n,\mm,\elll)\ \le\ 2^{O(k)} \left(
   n \frac{\elll}{w}\lg^{1/k}\mm\:\lg\left( \frac{w}{\elll} \lg\mm \right)
   \:+\: \mm M\right)
\]
for any value of $k$. Choosing $k=\sqrt{\lg\lg \mm}$ to asymptotically
minimize the expression, and plugging in $\elll=w$ and $M=\mm^2$ (so
that indeed $\frac{w}{\elll} \lg\mm \le \frac{1}{2}\lg M$), we get:
\[ T(n,\mm,w)\ =\ 2^{O(\sqrt{\lg\lg\mm})} \: (n  \:+\: \mm^3).
\]

We can reduce the dependence on $\mm$ to linear as follows. First,
select one out of every $\mm^{3/4}$ consecutive segments of $S$, and
run the above algorithm on just these $\mm^{1/4}$ segments. This takes
time $2^{O(\sqrt{\lg\lg\mm})} (n+\mm^{3/4})$ time. Now recurse
between each consecutive pair of selected segments.  The depth of the
recursion is $O(\lg\lg\mm)$, and it is straightforward to verify
that the running time is $n \cdot 2^{O(\sqrt{\lg\lg\mm})} + O(\mm)$.

\section{Avoiding Nonstandard Operations}  \label{sec:stdops}

The only nonstandard operation used by the algorithm of Section
\ref{sec:better} is applying a projective transform in parallel to
points packed in a word. Unfortunately, it does not seem possible to
implement this in constant time using standard word RAM operations
(since, according to the formula for projective transform, 
this operation requires multiple divisions where the divisors are all
different). 

One idea is to simulate the special operation in slightly
superconstant time. We can use the circuit simulation results of
Brodnik et al.~\cite{Bro} to reduce the operation to $\lg w \cdot
(\lg\lg w)^{O(1)}$ standard operations. 
For the version of the slab problem in dimension 3 or higher,
this is the best approach we know.  Note that the results from the previous
section hold in any constant dimension, by simply using the
multidimensional analog of Observation \ref{obs:basic} from
\cite{ChaPat}.

However, in two dimensions we can get rid of the dependence on the
universe, obtaining a time bound of $n\cdot 2^{O(\sqrt{\lg\lg m})}+O(m)$ on
the standard word RAM\@. This constitutes the object of this section.

\subsection{The Center Slab}

By horizontal translation, we can assume the left boundary of our
vertical slab is the $y$-axis. Let the abscissa of the right boundary
be $\Delta$. For some $h$ to be determined, let the \emph{center slab}
be the region of the plane defined by $\Delta / 2^h \le x \le \Delta
\cdot (1 - 2^{-h})$. The lateral slabs are defined in the intuitive
way: the left slab by $0 \le x \le \Delta / 2^h$ and the right slab by
$\Delta \cdot (1 - 2^{-h}) \le x \le \Delta$.

The key observation is that distances are somewhat well-behaved in the
center slab, so we will be able to decrease both the left and right
intervals at the same time, not just one of them. This enables us to
use (easier to implement) affine maps instead of projective
maps. Center slabs were also used in one of our previous papers
\cite{PatFOCS06}, but as presented there, the idea cannot get rid of
the dependence on the universe. This paper's definition and use of the
center slab is rather different, and gets rid of this dependence.

The following is a replacement for Observation \ref{obs:basic}:

\begin{observation}  \label{obs:center}
Fix $b$ and $h$. Let $S$ be a set of $m$ sorted disjoint segments,
such that all left endpoints lie on an interval $I_L$ and all right
endpoints lie on an interval $I_R$, where both $I_L$ and $I_R$ have
length $2^\ell$. In $O(b)$ time, we can find $O(b)$ segments $s_0,
s_1, \ldots \in S$ in sorted order, which include the lowest segment
of $S$, such that:
\begin{enumerate}
\item[\em (1)] for each $i$, at least one of the following holds:
  \begin{enumerate}
  \item[\em (1a)] there are at most $m/b$ segments of $S$ between
    $s_i$ and $s_{i+1}$.
  \item[\em (1b)] both the left and right endpoints of $s_i$ and
    $s_{i+1}$ are at distance at most $2^{\ell-h}$.
  \end{enumerate}

\item[\em (2)] there exist segments $\sss_0 \prec \sss_1 \prec \cdots$
  cutting across the slab, satisfying all of the following:
  \begin{enumerate}
  \item[\em (2a)] distances between the left endpoints of the
    $\sss_i$'s are multiples of $2^{\ell-2h}$.
  \item[\em (2b)] ditto for the right endpoints.
  \item[\em (2c)] inside the center slab, $s_0 \prec \sss_0 \prec s_2
    \prec \sss_2 \prec \cdots$.
  \end{enumerate}
\end{enumerate}
\end{observation}

\begin{figure}
\begin{center}
\setlength{\unitlength}{1.6in}
\input{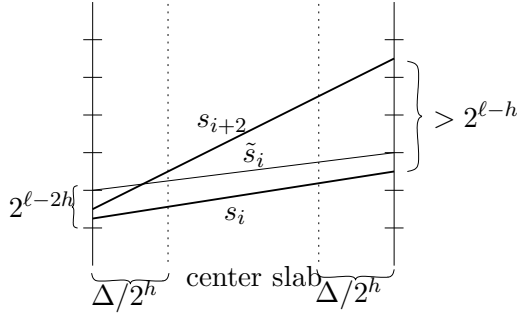}
\end{center}\vspace{-2.5ex}
\caption{Proof of Observation \protect\ref{obs:center}: a center slab.}
\label{fig:center}
\end{figure}

\begin{pf}
Let $B$ contain every $\down{m/b}$-th segment of $S$, starting with
the lowest segment $s_0$. We define $s_{i+1}$ inductively. If the next
segment after $s_i$ has either the left or right endpoints at distance
greater than $2^{\ell-h}$, let $s_{i+1}$ be this segment, which
satisfies (1a). Otherwise, let $s_{i+1}$ be the \emph{highest} segment
of $B$ which satisfies (1b).  

Now impose grids over $I_L$ and $I_R$, both consisting of $2^{2h}$
subintervals of length $2^{\ell-2h}$.  We obtain $\sss_i$ from $s_i$
by rounding each endpoint to the grid point immediately above.  This
immediately implies $\sss_0 \prec \sss_1 \prec \cdots$ and $s_i \prec
\sss_i$.  Unfortunately, $\sss_i$ and $s_{i+k}$ may intersect for
arbitrarily large $k$ (e.g.~$s_i \TO s_{i+k}$ are very close on the
left, while each consecutive pair is far on the right). However, we
will show that inside the center slab, $\sss_i \prec s_{i+2}$.
(See Figure~\ref{fig:center}.)

By construction, $s_i$ and $s_{i+2}$ are vertically separated by more
than $2^{\ell-h}$ either on the left or on the right. Since lateral
slabs have a fraction of $2^{-h}$ of the width of the entire slab, the
vertical separation exceeds $2^{\ell-h} / 2^h = 2^{\ell - 2h}$
anywhere in the center slab. Rounding $s_i$ to $\sss_i$ represents a
vertical shift of less than $2^{\ell-2h}$ anywhere in the slab. Hence,
$\sss_i \prec s_{i+2}$ in the center slab.
\end{pf}

\bigskip

We now describe how to implement $\slab()$, assuming the intervals
containing the left endpoints ($I_L$) and the right endpoints ($I_R$)
both have length $2^{\ell}$. In this section, we only deal with points
in the center slab. It is easy to $\Split()$ $Q$ into subsets
corresponding to the center and lateral slabs, and $\Unsplit()$ at the
end.

We use Observation \ref{obs:center} instead of Observation
\ref{obs:basic}. Since $I_L$ and $I_R$ have equal length, the map
$\varphi$ is affine. Thus, it can be implemented using parallel
multiplication and parallel addition. This means step~2 can be
implemented in time $O(n \frac{\elll}{w})$ using standard operations.

Because we only deal with points in the center slab, and there $s_i
\prec \sss_i \prec s_{i+2}$ (just like in the old Observation
\ref{obs:basic}), steps 4 and 5 work in the same way.

\subsection{Lateral Slabs}

To deal with the left and right slabs, we use the following simple
observation, which we only state for the left slab by symmetry. Note
that the guarantees of this observation (for the left slab) are
virtually identical to that of Observation \ref{obs:center} (for the
center slab). Thus, we can simply apply the algorithm of the previous
section for the left and right slabs.

\newcommand{\ttt}{\tilde{t}}

\begin{observation}  \label{obs:left}
Fix $b$ and $h$. Let $S$ be a set of $m$ sorted disjoint segments,
such that all left endpoints lie on an interval $I_L$ and all right
endpoints lie on an interval $I_R$, where both $I_L$ and $I_R$ have
length $2^\ell$. In $O(b)$ time, we can find $O(b)$ segments $t_0,
t_1, \ldots \in S$ in sorted order, which include the lowest segment
of $S$, such that:
\begin{enumerate}
\item[\em (1)] for each $i$, at least one of the following holds:
  \begin{enumerate}
  \item[\em (1a)] there are at most $m/b$ segments of $S$ between
    $s_i$ and $s_{i+1}$.
  \item[\em (1b)] anywhere in the left slab, the vertical separation
    between $s_i$ and $s_{i+1}$ is less than $2^{\ell-h+1}$.
  \end{enumerate}

\item[\em (2)] there exist segments $\ttt_0 \prec \ttt_1 \prec \cdots$
  cutting across the left slab, satisfying all of the following:
  \begin{enumerate}
  \item[\em (2a)] distances between the left endpoints of the
    $\ttt_i$'s are multiples of $2^{\ell-2h}$.
  \item[\em (2b)] ditto for the right endpoints.
  \item[\em (2c)] inside the left slab, $t_0 \prec \ttt_0 \prec t_2 \prec
    \ttt_2 \prec \cdots$.
  \end{enumerate}
\end{enumerate}
\end{observation}

\begin{figure}
\begin{center}
\setlength{\unitlength}{1.6in}
\input{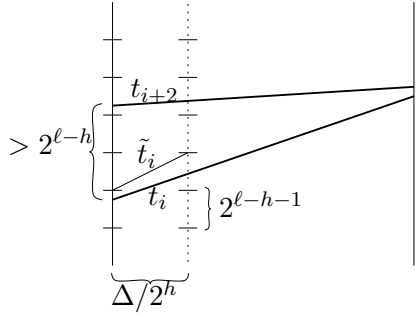}
\end{center}\vspace{-2.5ex}
\caption{Proof of Observation \protect\ref{obs:left}: a left slab.}
\label{fig:lateral}
\end{figure}

\begin{pf}
Let $I_A$ be the vertical interval at the intersection of the right
edge of the left slab with the parallelogram defined by $I_L$ and
$I_R$. Note $I_A$ also has size $2^\ell$.

Let $B$ contain every $\down{m/b}$-th segment of $S$, starting with
the lowest segment $t_0$. Given $t_i$, we define $t_{i+1}$ to be the
highest segment of $B$ which has the left endpoint at distance at most
$2^{\ell-h}$ away. If no such segment above $t_i$ exists, let
$t_{i+1}$ be the successor of $t_i$ in $B$ (this will satisfy
(1a)). In the first case, (1b) is satisfied because the right
endpoints of $t_i$ and $t_{i+1}$ are at distance most $2^\ell$, so on
$I_A$, the separation is at most $2^{\ell-h} (1 - 2^{-h}) + 2^{\ell}
\cdot 2^{-h} < 2^{\ell-h+1}$.

Now impose grids over $I_L$ and $I_A$, both consisting of $2^{h+1}$
subintervals of length $2^{\ell-h-1}$. We obtain $\ttt_i$ from $t_i$ by
rounding the points on $I_L$ and $I_A$ to the grid point immediately
above. Note that the vertical distance between $t_i$ and $\ttt_i$ is
less than $2^{\ell-h-1}$ anywhere in the left slab. On the other hand,
the left endpoints of $t_i$ and $t_{i+2}$ are at distance more than
$2^{\ell-h}$. The distance on $I_A$ (and anywhere in the left slab) is
at least $2^{\ell-h} (1-2^{-h}) \ge 2^{\ell-h-1}$. Thus $t_i \prec
\ttt_i \prec t_{i+2}$.  (See Figure~\ref{fig:lateral}.)
\end{pf}

\subsection{Bounding the Dependence on $m$}

Our analysis needs to be modified, 
because segments are simultaneously in the left, center and
right slabs, so they are included in 3 recursive calls.  In other
words, in recurrence (\ref{eq:new-rec}), we have to replace
$\sum_i m = m -b'$ with a weaker inequality $\sum_i m \le 3m$.
Recall that for our choice of $b$ and $h$, the depth of the
recursion is bounded by $O(\log_b \mm + \frac{\elll}{h})=
O(\lg^{1/(k+1)}\mm)$.
Thus, the cost per segment is increased by an 
extra factor of $3^{O(\lg^{(1/(k+1)}\mm)}=3^{O(\sqrt{\lg\mm})}$ 
for each bootstrapping round; the cost per point does not change.  
With $k=\sqrt{\lg\lg\mm}$ rounds, the overall dependence on $\mm$
is now increased slightly to
$2^{O(\sqrt{\lg\lg\mm})}\cdot \mm^3\cdot 3^{O(\sqrt{\lg\mm}\lg\lg\mm)})=
O(\mm^{3+\eps})$.
As before, this can be made $O(\mm)$ by working with
$\mm^{1/4}$ segments and recursing.


\section{Open Problems}

Though our algorithm for offline point location is deterministic, the
reductions to other problems \cite{ChaPat}
introduce randomization. It would be nice
to derandomize them. Another interesting question is 
whether some of
these reductions also hold in the other direction.  For example,
the trapezoidal decomposition problem of disjoint line
segments is clearly no easier than offline point location, but 
can the same be said for other problems like 3-d convex hulls?

A problem with an $O(n\lg n)$ upper bound that we currently cannot
improve is \emph{counting} intersections between a set of red segments
and a set of blue segments, given that no segments of the same color
intersect~\cite{ChEdGuALGMCA}.  Note that this problem is no easier
than counting inversions in a permutation, which takes $O(n \sqrt{\lg
  n})$ time by the best known algorithm~\cite{cp10invs}.

Finally, the complexity of the online point location problem
remains open: can the 
$O(\frac{\lg n}{\lg\lg n})$ and $O(\sqrt{\frac{w}{\lg w}})$ upper 
bounds from \cite{ChaPat} be improved, or can
stronger lower bounds be proved?

\small
\renewcommand{\baselinestretch}{0.98}

\bibliographystyle{abbrv}
\bibliography{nearlin}

\begin{thebibliography}{10}

\bibitem{AggVit}
A.~Aggarwal and J.~S. Vitter.
\newblock The input/output complexity of sorting and related problems.
\newblock {\em Commun. ACM}, 31(9):1116--1127, 1988.

\bibitem{AndFOCS96}
A.~Andersson.
\newblock Faster deterministic sorting and searching in linear space.
\newblock In {\em Proc. 37th IEEE Sympos. Found. Comput. Sci.}, pages 135--141,
  1996.

\bibitem{AndJCSS}
A.~Andersson, T.~Hagerup, S.~Nilsson, and R.~Raman.
\newblock Sorting in linear time?
\newblock {\em J. Comput. Sys. Sci.}, 57:74--93, 1998.

\bibitem{BeaFic}
P.~Beame and F.~Fich.
\newblock Optimal bounds for the predecessor problem and related problems.
\newblock {\em J. Comput. Sys. Sci.}, 65:38--72, 2002.

\bibitem{Bro}
A.~Brodnik, P.~B. Miltersen, and J.~I. Munro.
\newblock Trans-dichotomous algorithms without multiplication---some upper and
  lower bounds.
\newblock In {\em Proc. 5th Int. Workshop Algorithms Data Struct.}, pages
  426--439, London, UK, 1997. Springer-Verlag.

\bibitem{BucMul}
K.~Buchin and W.~Mulzer.
\newblock Delaunay triangulations in {$O(\mathrm{sort}(n))$} time and more.
\newblock In {\em Proc. 50th IEEE Sympos. Found. Comput. Sci.}, pages 139--148,
  2009.

\bibitem{cp10invs}
T.~M. Chan and M.~P{\v a}tra{\c s}cu.
\newblock Counting inversions, offline orthogonal range counting, and related
  problems.
\newblock In {\em Proc. 21st ACM/SIAM Sympos. Discrete Algorithms}, pages
  161--173, 2010.

\bibitem{ChaPat}
T.~M. Chan and M.~P\v{a}tra\c{s}cu.
\newblock Transdichotomous results in computational geometry, {I}: Point
  location in sublogarithmic time.
\newblock {\em SIAM J. Comput.}, 32(10):703--729, 2010.
\newblock Preliminary versions appeared in FOCS'06.

\bibitem{ChEdGuALGMCA}
B.~Chazelle, H.~Edelsbrunner, L.~J. Guibas, and M.~Sharir.
\newblock Algorithms for bichromatic line segment problems and polyhedral
  terrains.
\newblock {\em Algorithmica}, 11:116--132, 1994.

\bibitem{BerBOOK}
M.~de~Berg, M.~van Kreveld, M.~Overmars, and O.~Schwarzkopf.
\newblock {\em Computational Geometry: Algorithms and Applications}.
\newblock Springer-Verlag, Berlin, 1997.

\bibitem{EdeBOOK}
H.~Edelsbrunner.
\newblock {\em Algorithms in Combinatorial Geometry}, volume~10 of {\em EATCS
  Monographs on Theoretical Computer Science}.
\newblock Springer-Verlag, Heidelberg, West Germany, 1987.

\bibitem{HanINFCOMP}
Y.~Han.
\newblock Improved fast integer sorting in linear space.
\newblock {\em Inf. Comput.}, 170:81--94, 2001.

\bibitem{HanSTOC02}
Y.~Han.
\newblock Deterministic sorting in {$O(n\log\log n)$} time and linear space.
\newblock In {\em Proc. 34th ACM Sympos. Theory Comput.}, pages 602--608, 2002.

\bibitem{HanThoFOCS02}
Y.~Han and M.~Thorup.
\newblock Integer sorting in {$O(n\sqrt{\log\log n})$} expected time and linear
  space.
\newblock In {\em Proc. 43rd IEEE Sympos. Found. Comput. Sci.}, pages 135--144,
  2002.

\bibitem{KirRei}
D.~G. Kirkpatrick and S.~Reisch.
\newblock Upper bounds for sorting integers on random access machines.
\newblock {\em Theoret. Comput. Sci.}, 28:263--276, 1984.

\bibitem{MulBOOK}
K.~Mulmuley.
\newblock {\em Computational Geometry: An Introduction Through Randomized
  Algorithms}.
\newblock Prentice Hall, Englewood Cliffs, NJ, 1994.

\bibitem{OroBOOK}
J.~O'Rourke.
\newblock {\em Computational Geometry in {C}}.
\newblock Cambridge University Press, 2nd edition, 1998.

\bibitem{PreSha}
F.~P. Preparata and M.~I. Shamos.
\newblock {\em Computational Geometry: An Introduction}.
\newblock Springer-Verlag, New York, NY, 1985.

\bibitem{PatFOCS06}
M.~P\v{a}tra\c{s}cu.
\newblock Planar point location in sublogarithmic time.
\newblock In {\em Proc. 47th IEEE Sympos. Found. Comput. Sci.}, pages 325--332,
  2006.

\bibitem{SnoSURV}
J.~Snoeyink.
\newblock Point location.
\newblock In J.~E. Goodman and J.~O'Rourke, editors, {\em Handbook of Discrete
  and Computational Geometry}, pages 767--785. CRC Press LLC, Boca Raton, FL,
  2nd edition, 2004.

\bibitem{ThoSODA97}
M.~Thorup.
\newblock Randomized sorting in {$O(n\log\log n)$} time and linear space using
  addition, shift, and bit-wise boolean operations.
\newblock {\em J. Algorithms}, 42:205--230, 2002.

\end{thebibliography}

\end{document}